\documentstyle[aps,epsf,preprint,graphicx,prl]{revtex}
\begin{document} 
\draft
\newcounter{saveeqn}
\newcommand{\alpheqn}{\setcounter{saveeqn}{\value{equation}}%
  \stepcounter{saveeqn}\setcounter{equation}{0}%
  \renewcommand{\theequation}{%
  \mbox{\arabic{saveeqn}\alph{equation}}}}%
\newcommand{\reseteqn}{\setcounter{equation}{\value{saveeqn}}%
\renewcommand{\theequation}{\arabic{equation}}}
\def\beq{\begin{equation}}
\def\eeq{\end{equation}}
\def\beqn{\begin{eqnarray}}
\def\eeqn{\end{eqnarray}}

\title{Interferences in the density of two initially independent Bose-Einstein condensates}

\author{L. S. Cederbaum$^{1}$, A. I. Streltsov$^{1}$, Y. B. Band$^{2}$, and O. E. Alon$^{1}$}
\address{$^1$Theoretical Chemistry, 
Heidelberg University, Im Neuenheimer Feld 229, 69120 Heidelberg \&
Max-Planck Institute for Nuclear Physics, Saupfercheckweg 1, 69117 Heidelberg, Germany}
\address{$^2$Department of Chemistry, Ben-Gurion University of the Negev,
Beer-Sheva 84105, Israel}

\maketitle

\begin{abstract}
It is shown that the density of two {\it initially independent} condensates which
are allowed to expand and overlap can show interferences as a function of time
due to interparticle interaction.
Using many-body theory,
explicit expressions for the density are given which are
exact in the weak interaction limit.
General working equations are discussed which reproduce exactly the density in this limit.
Illustrative examples are presented.
\end{abstract}

The investigation of interferences between particles is one of the
most basic tools to learn on the nature of quantum gases.
Interferences attracted much attention in particular in the case of
Bose-Einstein condensates (BECs) both from the theoretical and experimental sides,
see, e.g., \cite{Ketterle,Schmid,Javanainen,Wallis0,Castin,Chem_phys,Leggett}.
In a popular set up studied,
identical atoms are produced in two
traps which we may call the left and right traps and which are separated by a barrier.
By removing the traps and the barrier between them,
the atoms expand freely and can overlap.
In experiment, the photographs obtained show
spectacular interference fringes \cite{Ketterle,Schmid}.

The interference of two parts of a single coherent condensate is by now well understood,
see, e.g., \cite{Wallis0,Wallis1,Niu}.
On the other hand, very little is known on the interference of two
initially independent (i.e., fragmented) BECs, except for the case 
of non-interacting particles \cite{Javanainen,Castin,Chem_phys}.
Fragmented BECs can be produced using a barrier between the two traps 
which is so high and broad that tunneling between them is negligible.

In the available experiments, the atoms are prepared in a double-well trap
potential and it is not generally proven whether the atoms form a coherent
BEC, a fragmented BEC, or a combination thereof.  Moreover, it is feasible
nowadays to produce in the lab two spatially separate, initially independent
BECs, see, e.g., \cite{Shin_Frag}, and this allows for experiments with 
definitely fragmented BECs.
Apart from its importance as a fundamental problem, 
the solution of the problem of interference of two initially independent condensates 
is thus also of practical relevance.

It is obvious from the literature that the impact of interaction on
the interferences between two initially independent BECs is not clarified. 
This is documented, e.g., in the review \cite{Leggett}.
Can at all interactions between two initially independent BECs lead 
to visible interferences in the density? 
If yes, how do they look like? 
On what do they depend? 
These are the questions to which we provide clear and novel answers.

In the scenario of two initially independent BECs the initial state 
of the many-body system before removing the traps reads
\beq\label{eq1}
 \left|\Psi\right> = \left(N_L! N_R!\right)^{-1/2} 
{(b_L^\dag)}^{N_L}{(b_R^\dag)}^{N_R} \left|vac\right>,
\ \ N_L+N_R=N, 
\eeq
where the $b_L^\dag$ and $b_R^\dag$ are the usual creation operators
for bosons in the left and right traps, respectively,
which contain definite numbers $N_L$ and $N_R$ of atoms in them.
After removing the traps, 
the state $\left|\Psi\right>$
is no longer an eigenstate of the system's Hamiltonian $H_0$
and expands in space as a function of time.
The time-dependent density, i.e., the expectation value of the density operator $\hat\rho(x)$
as a function of time becomes \cite{Leggett} 
\beq\label{eq2}
 \rho(x,t) \equiv \left<\Psi(t)\left|\hat\rho(x)\right|\Psi(t)\right> =
 N_L\left|\Phi_L(x,t)\right|^2 + N_R\left|\Phi_R(x,t)\right|^2, 
\eeq
where the $\Phi_{L,R}(x,t)$ are the single-atom states corresponding to\break\hfill
$b_{L,R}(t)=\exp(iH_0t)b_{L,R}\exp(-iH_0t)$.
Obviously, the density is a sum of the individual densities of the
two condensates and does not exhibit an interference term.

We would like to draw attention to the fact that the literature result 
(\ref{eq2}) has been obtained under the assumption that atoms belonging
to the two different BECs do not interact with each other.
In the following we demonstrate that in the presence of interaction,
the density $\rho(x,t)$ does show an interference term.
This finding has many consequences.
In particular, the corresponding interference structures remain
after the statistical averaging over many experimental runs.
Of course, as $\rho(x,t)$ changes in time, 
the average must be carried out at the same value of $t$.

The density operator $\hat\rho(x)=\hat\Psi^\dag(x)\hat\Psi(x)$,
where $\hat\Psi(x)$ is the usual field operator,
can be expressed in any complete basis set of one-particle functions.
Using (box normalized) plane waves, one has
\beq\label{eq3}
 \hat\rho(x) = v^{-1} \sum_{k,k'} e^{i(k-k')x} a_{k'}^\dag a_k,
\eeq
which is a suitable choice for our scenario.
Here, $v$ is the volume and $a_k$ is the destruction operator of a free boson
with momentum $k$ having the usual commutation relations
$[a_k,a^\dag_{k'}]=\delta_{kk'}$.
In this representation, the Hamiltonian of interacting identical bosons
of mass $m$, after release of the traps, is given by
\beqn\label{eq4}
& & H=H_0+V, \ \ \ \ H_0=\sum_k \frac{k^2}{2m} a_k^\dag a_k, \nonumber \\
& & V=\frac{\lambda_a}{2v} \sum_{k_1,k_2,k_3,k_4} \delta_{k_1+k_2,k_3+k_4} 
 a_{k_1}^\dag a_{k_2}^\dag a_{k_3} a_{k_4}. \ 
\eeqn
For the ease of presentation, the widely used contact interaction 
$V(x,x')=\lambda_a\delta(x-x')$,
where $\lambda_a$ is proportional to the s-wave scattering length, is used \cite{Leggett,Stringari_book}.
Of course, any other interparticle interaction can be used as well.
As usual, $H_0$ describes the motion of the free atoms.

We may now proceed to evaluate the time-dependent density $\rho(x,t)$
of the interacting particles.
First we transform to the interaction picture and write
$\hat\rho_I(x,t)=e^{iH_0t}\hat\rho(x)e^{-iH_0t}$ and all other
quantities in this picture are extremely easy to evaluate.
In particular, 
$a_k(t)\equiv e^{iH_0t} a_k e^{-iH_0t} = e^{-i\frac{k^2}{2m}t}a_k$.
To proceed we make use of the identity 
\alpheqn
\beq\label{eq5a}
 \rho(x,t)=\left<\Psi_I(t)\left|\hat\rho_I(x,t)\right|\Psi_I(t)\right> 
\eeq
where $\left|\Psi_I(t)\right>=e^{iH_0t}e^{-iHt}\left|\Psi(0)\right>$.
This expression for the density can be systematically evaluated 
using the text-book expansion
\beq\label{eq5b}
 \left|\Psi_I(t)\right> = \left\{1-i\int_0^t V_I(t_1)dt_1 
+ (-i)^2 \int_0^t V_I(t_1)dt_1 \int_0^{t_1} V_I(t_2)dt_2 + \ldots \right\} \left|\Psi(0)\right>.  
\eeq
\reseteqn
Inserting this expression into the expression (\ref{eq5a})
for $\rho(x,t)$ gives
\beqn\label{eq6}
 \rho(x,t)&=&\rho^0(x,t) + i\int_0^t dt_1 
 \left<\Psi\left|\left[V_I(t_1),\hat\rho_I(x,t)\right]\right|\Psi\right>  \nonumber \\
 &+& (i)^2 \int_0^t dt_1 \int_0^t dt_2 
 \left<\Psi\left|\left[V_I(t_2),\left[V_I(t_1),\hat\rho_I(x,t)\right]\right]\right|\Psi\right>
 + \ldots \
\eeqn
which is a systematic expansion of $\rho(x,t)$ in terms of the interparticle interaction.
$\left|\Psi\right>\equiv\left|\Psi(0)\right>$ is the initial state in Eq.~(\ref{eq1}),
which is an eigenstate of the system in the presence of the trap potentials,
and $\rho^0(x,t)=\left<\Psi\left|\hat\rho_I(x,t)\right|\Psi\right>$ is the 
{\it free density} evolving without the impact of this interaction.
With (\ref{eq6}) and the commutator relations for the boson operators 
$a_k$ and $a^\dag_{k'}$,
one obtains an explicit expansion of $\rho(x,t)$.
With growing order, the terms of this expansion
contain higher products of destruction and creation operators,
$a^\dag_{k_1}a_{k_2}$, $a_{k_1}^\dag a_{k_2}^\dag a_{k_3} a_{k_4}$, and so on.

How can one now evaluate the desired density $\rho(x,t)$
in view that the initial state $\left|\Psi\right>$ in Eq.~(\ref{eq1}) is
expressed in terms of other operators and not in terms of $a_k^\dag$?
To this end we remind that the field operator $\hat\Psi$ can be expressed in any 
complete set of functions,
plane waves or others.
Since $b_L$ and $b_R$ obviously commute,
the respective functions $\Phi_L$ and $\Phi_R$ may be seen as the first two
members in the expansion of the field operator:
$\hat\Psi(x)=\Phi_L(x)b_L+\Phi_R(x)b_R+{\mathit{''Rest\,''}}$.
It immediately follows that
\beq\label{eq7}
a_k=\varphi_L(k)b_L+\varphi_R(k)b_R+{\mathit{''Rest(k)\,''}},
\eeq
where $\varphi(k)$ is the Fourier transform of $\Phi(x)$
and the ${\mathit{''Rest\,''}}$ contains all other destruction operators commuting with $b_L$ and $b_R$.
Luckily, all the latter operators have a vanishing impact when 
applied to the initial state $\left|\Psi\right>$.
Consequently, if we order all the creation operators to the left and all the
annihilation operators to the right when
taking the expectation
value of $\hat\rho(x,t)$ with the many-body state $\left|\Psi\right>$ in Eq.~(\ref{eq1}),
we can throw the ${\mathit{''Rest\,''}}$ out in the following.
Inserting the result (\ref{eq7}) into the above mentioned higher products of
destruction and annihilation operators in the expansion of $\hat\rho(x,t)$,
one readily obtains
\beq\label{eq8}
 \rho(x,t)=\rho_{LL}(x,t)+\rho_{RR}(x,t)+\rho_{LR}(x,t),
\eeq
where $\rho_{LL}$ and $\rho_{RR}$ are the densities of the 
expanding separated BECs as if the two BECs do not communicate,
and $\rho_{LR}$ is the change of the density due to the interaction between them.
The terms contributing to $\rho_{LL}$ ($\rho_{RR}$) contain only
$b_L (b_R)$ and $b^\dag_L (b^\dag_R)$ operators, e.g.,
$b^\dag_L b^\dag_L b_L b_L$,
and those contributing to $\rho_{LR}$ contain only mixed products, e.g., $b^\dag_L b^\dag_R b_L b_R$.

Note that the calculation of $\rho(x,t)$ for interacting
bosons amounts to solving a full many-body problem.
Wishing to arrive at an exact analytical result, 
we concentrate first on weak interparticle interactions.

With $V$ in Eq.~(\ref{eq4}) and $\hat\rho(x)$ in Eq.~(\ref{eq3}) 
and the very simple appearance of $a_k(t)$ given above,
we evaluated explicitly the commutator $\left[V_I(t_1),\hat\rho_I(x,t)\right]$
and with it the leading term in Eq.~(\ref{eq6}).
The calculation is somewhat lengthy,
but very straightforward.
Using the simple ``trick'' (\ref{eq7}) we obtain the {\it exact}
results up to first order 
\alpheqn
\beqn\label{eq9a}
& & \rho_{LR}=-4\lambda_a N_L N_R \,
  {\mathrm Im}\!\left\{\Phi_L(x,t)A_{LR}(x,t)+\Phi_R(x,t)A_{RL}(x,t)\right\}, \nonumber \\
& & \rho_{LL}= N_L\left|\Phi_L(x,t)\right|^2 - 2\lambda_a N_L(N_L-1)
  {\mathrm Im}\!\left\{\Phi_L(x,t)A_{LL}(x,t)\right\}. \
\eeqn
Here $\Phi_{L,R}(x,t)$ is the freely expanding $\Phi_{L,R}(x)$ and
the amplitude $A_{LR}$ reads
\beq\label{eq9b}
A_{LR} = \frac{1}{v^{3/2}} \sum_{k_1,k_2,k_3} e^{i(k_3-k_1-k_2)x}
\varphi^\ast_L(k_1)\varphi^\ast_R(k_2)\varphi_R(k_3)
 \int_0^t dt_1 e^{i[(k_1^2+k_2^2-k_3^2)t_1+(k_1+k_2-k_3)^2(t-t_1)]/(2m)}. 
\eeq
\reseteqn
To obtain $\rho_{RR}$ just interchange $L$ and $R$ in (\ref{eq9a}).
The quantities $A_{RL}$, $A_{LL}$ and $A_{RR}$
are obtained analogously from (\ref{eq9b}).
The integration over $t_1$ in Eq.~(\ref{eq9b}) can, of course, be performed
explicitly, but then the summations over the three momenta
are more cumbersome to carry out.
Let us briefly discuss the result (9).
Clearly, the interference term $\rho_{LR}$ vanishes for $t\to 0$.
Furthermore, $\rho_{LR}(x,t)$ vanishes as expected if the atoms do not interact with
each other ($\lambda_a \to 0$).
The interference term $\rho_{LR}(x,t)$ is enhanced by the product $N_LN_R$
of the numbers of atoms in the two initial BECs.

The above results make clear that the interaction between the particles
gives rise to an interference term in the density of
two initially independent BECs of identical bosons.
Before presenting a numerical example we go one step further
and pose the question whether we can formulate a mean-field
theory which reproduces exactly the exact small $\lambda_a$ result (\ref{eq9a}).
Such a theory would open the door for real applications.
The standard mean-field leads to the well-known and widely used
Gross-Pitaevskii equation \cite{Leggett,Stringari_book}.
For coherent states this equation gives exact results for small $\lambda_a$.
Clearly, it is inapplicable to fragmented states (\ref{eq1}).
For fragmented states a more general multi-orbital
mean-field theory has been recently derived \cite{frag0}.
In the present scenario two orbitals are involved and the respective
time-dependent mean-field (TDMF($2$)) takes on the appearance
(for the general derivation of TDMF, see \cite{TDMF}):
\beqn\label{eq10}
 i \dot\psi_L = {\mathcal P} \left[\hat h + \lambda_a(N_L-1)\left|\psi_L\right|^2 +
   2 \lambda_aN_R\left|\psi_R\right|^2\right] \psi_L, \nonumber \\
 i \dot\psi_R = {\mathcal P} \left[\hat h + \lambda_a(N_R-1)\left|\psi_R\right|^2 +
   2 \lambda_aN_L\left|\psi_L\right|^2\right] \psi_R \
\eeqn
where the initial conditions are 
$\psi_{L,R}(x,t=0)=\Phi_{L,R}(x)$.
$\hat h$ is the usual one-particle Hamiltonian
(in our scenario just the kinetic energy operator) and
${\mathcal P}=1-\left|\psi_L\left>\right<\psi_L\right|-\left|\psi_R\left>\right<\psi_R\right|$
is a projector which ensures orthonormalization of the orbitals $\psi_L$ and $\psi_R$ \cite{TDMF}.

We now prove that the TDMF($2$) in (\ref{eq10}) exactly reproduces
the exact many-body small $\lambda_a$ result (\ref{eq9a}).
In TDMF($2$) the density can be expresses by 
$\rho(x,t)=N_L\left|\psi_L(x,t)\right|^2+N_R\left|\psi_R(x,t)\right|^2$.
Since $\lambda_a$ in (\ref{eq9a}) is taken to be small,
we may write $\psi_L=\Phi_L(x,t)+\delta\psi_L(x,t)$ and analogously for $\psi_R$.
Inserting into the latter expression for the density and comparing with (\ref{eq9a}),
we immediately identify $\delta\psi_L$:
\beq\label{eq11}
\delta\psi_L = -i\lambda_a(N_L-1)A^\ast_{LL}-i2\lambda_aN_RA^\ast_{LR}.
\eeq
To obtain $\delta\psi_R$ just interchange $L$ and $R$.
Of course, we still have to show that $\psi_L=\Phi_L+\delta\psi_L$ with
$\delta\psi_L$ from Eq.~(\ref{eq11}) indeed fulfills the TDMF equations,
i.e., is the solution of (\ref{eq10}).
Taking the derivative of $\Phi_L+\delta\psi_L$ with respect to time
and using $i\dot\Phi_L=\hat h\Phi_L$, leads to
\alpheqn
\beq\label{eq12a}
i\dot\psi_L = \hat h \Phi_L + \lambda_a(N_L-1)\dot A^\ast_{LL} + 2\lambda_a N_R \dot A^\ast_{LR} \ .
\eeq
$\dot A^\ast_{LL}$ and $\dot A^\ast_{LR}$ can be deduced from Eq.~(\ref{eq9b}).
The time $t$ appears there twice.
The derivative with respect to the upper limit
of the integral leaves us with a separable triple sum of terms
like $v^{-1/2} \sum e^{ik_1x} e^{ik_1^2/(2m)} \varphi_L(k_1)=\Phi_L(x,t)$,
which leads to $\left|\Phi_L\right|^2\Phi_L$ and $\left|\Phi_R\right|^2\Phi_L$, respectively.
The derivative with respect to $t$ in the exponential function gives an expression
which is identical to $-\hat h A^\ast_{LL}$ and $-\hat h A^\ast_{LR}$, respectively
(remember that $\hat h = -\frac{1}{2m}\frac{\partial^2}{\partial x^2}$).
Collecting all terms yields
\beq\label{eq12b}
i\dot\psi_L = \hat h \psi_L + \left[\lambda_a(N_L-1)\left|\Phi_{L}\right|^2+
 2 \lambda_aN_R\left|\Phi_R\right|^2\right] \Phi_L \
\eeq
\reseteqn
which is, as the orthonormalization is of second order in $\lambda_a$,
identical with the TDMF($2$) in Eq.~(\ref{eq10}) to first order
in the interaction strength $\lambda_a$.

In the following we present two illustrative numerical examples.
In the first the interaction is very weak and
the initial single-atom functions $\Phi_{L,R}(x)$ are normalized
Gaussians located at $\pm x_0$, i.e.,
$\Phi_{L,R}(x)=(\frac{2}{\pi a^2})^{1/4} \exp\{-(x\mp x_0)^2/a^2\}$.
This is a realistic choice for harmonic traps.
The Fourier transforms of these functions simply are
$\varphi_{L,R}(k)=a^{1/2}/(2\pi)^{3/4}\exp(-k^2a^2/4)\exp(\mp ikx_0)$.
They evolve in time as 
$\varphi_{L,R}(k,t)=e^{-i\frac{k^2}{2m}t}\varphi_{L,R}(k)$ and their
back-transforms give $\Phi_{L,R}(x,t)$ which are normalized Gaussians expanding in space
as a function of time and can be found in many elementary text books.
For the ease of presentation our example is in one dimension.
After transferring the summations in Eq.~(\ref{eq9b}) to integrals
over the momenta,
all three integrals over $k_3$, $k_2$ and $k_1$ can be carried out analytically 
using that $\int_{-\infty}^{+\infty} e^{-q^2(k+p)^2} dk = \sqrt{\pi}/q$
for any complex quantities $q,p$ as long as ${\mathrm Re} \, q^2 > 0$.
We have checked that this condition is fulfilled for all three integrals.
The final result is of the form
$A_{LR}(x,t)=\int_0^t dt_1 f(t_1)e^{g(x,x_0,t_1,t)}$,
where $f(t_1)$ and $g(x,x_0,t_1,t)$ are simple but lengthy algebraic expressions.

Fig.~1 shows the interference term of the density $\rho_{LR}(x,t)$ 
for very weak atom-atom interaction.
Without interaction $\rho_{LR}(x,t)=0$
and the density is simply 
$N_L\left|\Phi_L(x,t)\right|^2 + N_R\left|\Phi_R(x,t)\right|^2$.
The interaction leads to $\rho_{LR}$ for which
we use expression (9).
To simplify the discussion, we put $m=a/2=\hbar=1$ and 
express $x$ in units of $a/2$,
$t$ in units of $(a/2)^2 m$,
and $\lambda_{a}$ in units of $\frac{1}{(a/2)^3m}$.
Furthermore, $N_L=N_R=500$, $x_0=6$ and $\lambda_{a}=2.5 \cdot 10^{-7}$.
At $t=0$ one has $\rho_{LR}(x,0)=0$ and then it starts to grow as $\sim t^2$ at very short times
and much faster later on.
$\rho_{LR}(x,t)$ exhibits an oscillatory behavior which changes as time proceeds.

In the following we apply the TDMF theory.
As a first step we compute $\rho_{LR}(x,t)$ for the case
of weak interaction discussed above.
The results are also depicted in Fig.~1 and seen to coincide with those obtained
using the analytic expression (9).
Next, we enlarge the interaction strength $\lambda_a$.
For coherent states the time-dependent Gross-Pitaevskii equation,
which is exact in the weak interaction limit,
has been demonstrated in many cases to be applicable for intermediate
and stronger interactions \cite{Leggett,Stringari_book}.
Similarly, there is reason to expect that for fragmented 
states the TDMF theory, which has been proven above to be exact in the
weak interaction limit, is applicable well beyond this limit.
We mention that TDMF($1$) is nothing but the 
time-dependent Gross-Pitaevskii equation.

For the very weak interaction discussed above we have chosen
as initial conditions Gaussians located at $\pm x_0$.
Now, as the interaction is increased to $\lambda_a=0.1$, 
we choose the respective
solutions of the stationary Gross-Pitaevskii equation at this $\lambda_a$
as initial conditions to account for the interaction when the harmonic traps
centered at $\pm x_0$ are released.
In Fig.~2 the density $\rho(x,t)$ computed using the TDMF($2$) equations (\ref{eq10})
is shown as a function of time.  
As seen in the figure, 
at $t=0$ the density consists of two separated distributions centered at $\pm x_0$.
The traps are removed at this time and the distributions start to broaden and to overlap.
At about $t=2.4$ one begins to see impact of the interference term in the density which
becomes strongly pronounced as time proceeds.

We conclude that the density of two initially independent condensates which are
allowed to overlap can show interference effects in the presence of interparticle interaction.
The physics of so called fragmented states, like the state in Eq.~(\ref{eq1}),
is generally very different from that of coherent states \cite{frag1,frag2}.
Coherent states of condensates have been extensively studied,
mostly in the framework of the Gross-Pitaevskii equation \cite{Leggett,Stringari_book}.
A BEC in a coherent state can exhibit interference fringes
even in the absence of interaction \cite{Leggett,Wallis1,Niu,Pitaevskii,markus}.
Take, for instance, the coherent state
$\left|\Psi^{coh}\right> = (N!)^{-1/2}{(b^\dag)}^{N}\left|vac\right>$
with $b^\dag=(b^\dag_L+b^\dag_R)/\sqrt{2}$.
This immediately leads to
$\rho^{coh}(x,t)=\frac{N}{2}\left|\Phi_L(x,t)+\Phi_R(x,t)\right|^2$ in the
absence of interaction between the atoms.
In analogy to Eq.~(\ref{eq7}) we can determine the interference term
$\rho_{LR}^{coh}(x,t)=N{\mathrm Re}\left(\Phi_L^\ast\Phi_R\right)$.
For the expanding Gaussians discussed above,
the oscillatory part of $\rho_{LR}^{coh}$ is simply given by
$\cos[K(t)x]$ with $K(t)=8x_0(t/m)/(a^4+4t^2/m^2)$.
This interference term is qualitatively different from that arising due to the 
interaction between the particles.
Another important difference between $\rho^{coh}_{LR}(x,t)$ and $\rho_{LR}(x,t)$
worth mentioning is that the former depends on the relative phase
between $\Phi_L(x)$ and $\Phi_R(x)$, 
while the latter does not depend on this phase.

Whether in an experiment the initial state is coherent or fragmented depends on the 
experimental conditions. 
It is beyond the scope of this work to argue whether or not the
initial state in the currently available experiments on interference is fragmented.
It is also not our intention to take side in the ongoing debate on whether these 
experiments detect the density or higher-order
correlation functions, 
although we tend to share the opinion of some researchers
see, e.g., \cite{Wallis0,Wallis1,Niu,Pitaevskii},
that the density is measured.
What we can state, is that if one
measures the density of two {\it freely} expanding initially independent BECs,
it will only show interferences in the presence of interaction.
This leads to the following proposal for an experiment
which makes use of the fact that nowadays one can vary the strength
of the interaction between the atoms \cite{Feshbach_resonance1,Feshbach_resonance2}.
Two measurements are necessary.
If the measurement with interaction shows interferences which
disappear upon measuring with the interaction turned off,
then (a) the initial state was a fragmented state 
and (b) the interaction is responsible for the interferences.

The theory presented here is easily extendable to any kind of interparticle interaction. 
It is also easily extendable to the case where one does not let the
two BECs expand freely by removing the traps completely.
One may, e.g., remove only the barrier
and let the BECs expand in the new global trap.
Since the interference structures depend on the interaction,
a wealth of effects can be expected by varying the interaction,
the form of the individual traps
and of the numbers $N_L$ and $N_R$ of the particles.

\acknowledgments

\noindent
Financial support and collaboration within the 
German-Israeli DIP project of the German Federal Ministry of Education and Research
(BMBF) are greatfully acknowledged.

\begin{figure}[ht] 
\includegraphics[width=11cm,angle=-90]{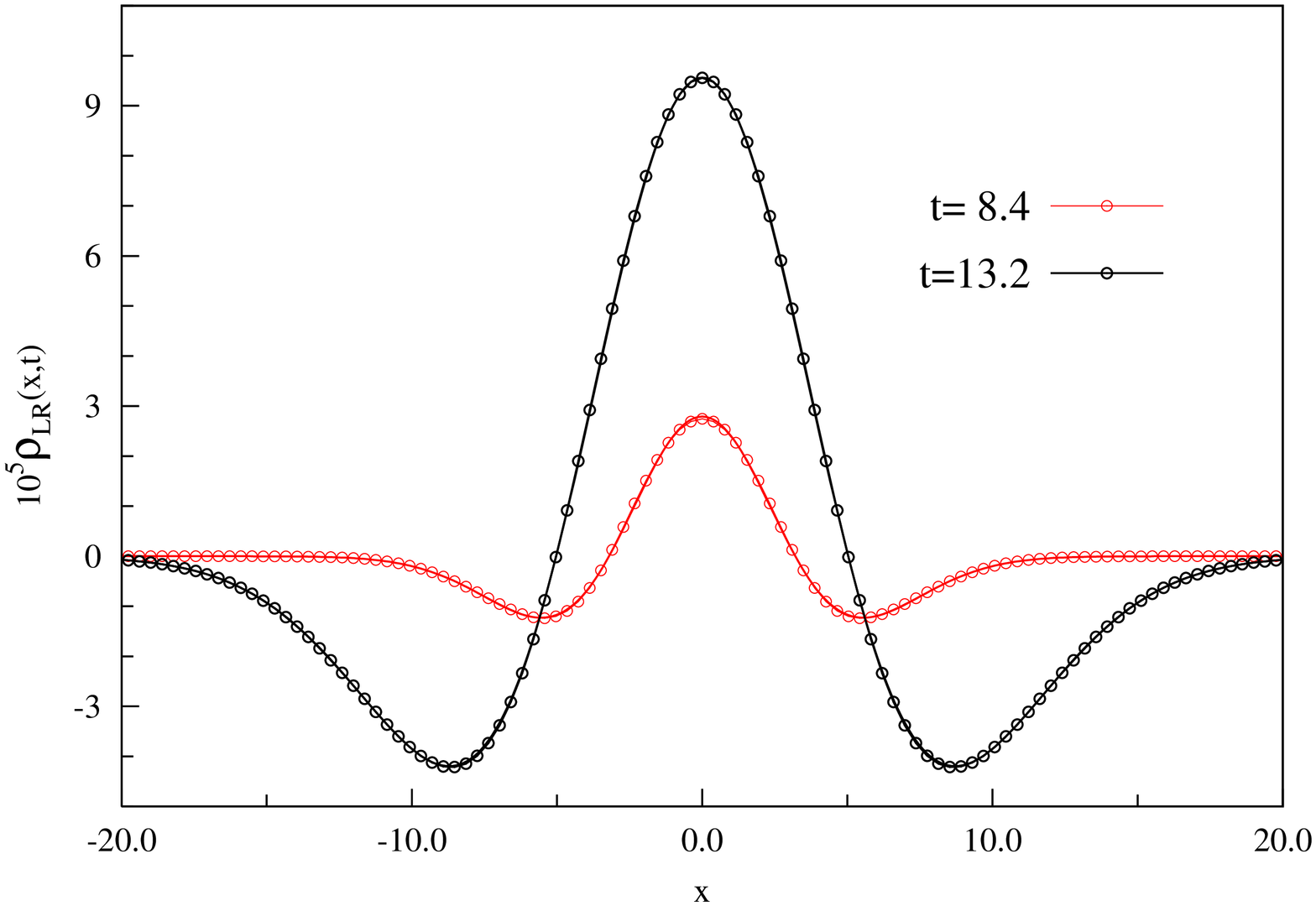}
\vglue 1.0 truecm
\caption [kdv] {(Color online) The interference term $\rho_{LR}$ for very weak interaction 
strength ($\lambda_{a}=2.5 \cdot 10^{-7}$) at two values of $t$.
The solid curves are computed using the analytic result (9).
The dots are computed using the TDMF($2$) equations (\ref{eq10}).
The quantities shown are dimensionless.}
\end{figure}

\begin{figure}[ht] 
\includegraphics[width=11cm,angle=-0]{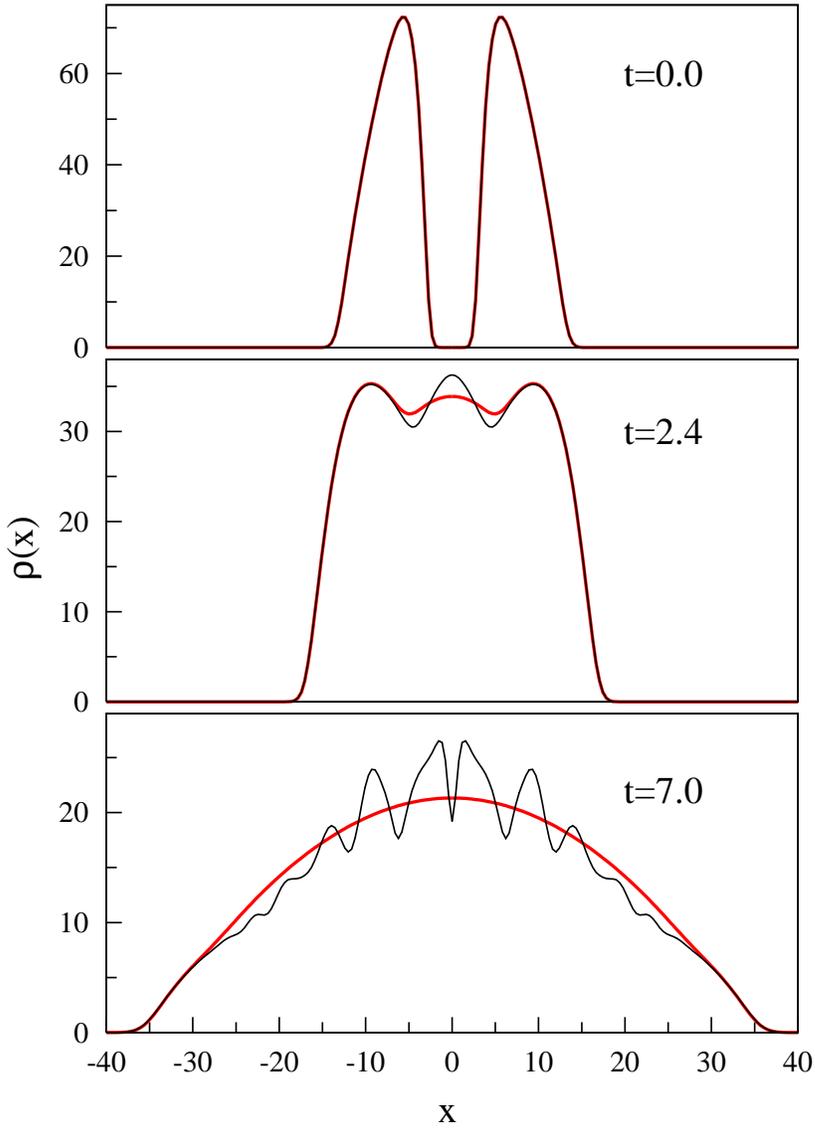}
\vglue 1.0 truecm
\caption [kdv]{(Color online) The density $\rho(x,t)$ of two condensates
of $500$ atoms each for $\lambda_{a}=0.1$ as a function of time computed with TDMF($2$) (black)
compared to the density $\rho_{LL}+\rho_{RR}$ of
two BECs which do not interact with each other,
each computed with the Gross-Pitaevskii equation (red).
The quantities shown are dimensionless.
For more details see text.}
\end{figure}

\end{document}